\documentclass{PoS}
\usepackage{amsmath}
\usepackage{amssymb}

\title{Lattice Planar QED in external magnetic field}

\ShortTitle{Lattice Planar QED in external magnetic field}

\author{Paolo Cea\\
        Dipartimento di Fisica dell'Universit\`a di Bari and \\
        INFN, Sezione di Bari, Italy\\
        E-mail: \email{paolo.cea@ba.infn.it}}

\author{Leonardo Cosmai\\
        INFN, Sezione di Bari, Italy\\
        E-mail: \email{leonardo.cosmai@ba.infn.it}}

\author{\speaker{Pietro Giudice}\\
         Department of Physics, College of Science, 
         Swansea University, United Kingdom\\
        E-mail: \email{p.giudice@swansea.ac.uk}}

\author{Alessandro Papa\\
        Dipartimento di Fisica dell'Universit\`a della Calabria and \\
        INFN, Gruppo collegato di Cosenza, Italy \\
        E-mail: \email{papa@cs.infn.it}}

\abstract
{
We investigate planar Quantum ElectroDynamics (QED) with two degenerate
staggered fermions in an external magnetic field on the lattice. Our
preliminary results indicate  that in external magnetic fields there is
dynamical generation of mass for two-dimensional massless Dirac fermions in
the weak coupling region. We comment on possible implications to the quantum
Hall effect in graphene.
}

\FullConference{XXIX International Symposium on Lattice Field Theory \\
                 July 10 - 16, 2011\\
                 Squaw Valley, Lake Tahoe, California}

\newcommand{\beq}{\begin{equation}}
\newcommand{\eeq}{\end{equation}}
\newcommand{\bea}{\begin{eqnarray}}
\newcommand{\eea}{\end{eqnarray}}

\begin{document}

\section{Introduction}

Understanding the properties of the vacuum of gauge theories is still an 
interesting field, potentially rich of surprises and unexpected applications.

It is well recognised today that relativistic field models can
be used as effective theories to describe the low-energy excitations in 
condensed matter systems; in particular, they can be applied to a wide class 
of planar systems (see for example Ref.~\cite{Zee:2003mt}, part VI).

Recently much attention has been devoted to the application
of the lattice approach to describe some properties of a
single sheet of carbon atoms arranged in the well known 
honeycomb structure called ``graphene'' (among the first papers that 
appeared we want to remind Ref.~\cite{Hands:2008id} and 
Ref.~\cite{Drut:2008rg}).

The vacuum structure of lattice gauge theories can be understood 
probing it by an external background field $\vec{A}^{\text{ext}}$.
In this paper, we show that the presence of an external magnetic field induces
a dynamical symmetry breaking in the particular case of QED in $2+1$ 
dimensions with dynamical fermions.

We will also discuss how our results can be applied to describe some
striking properties, observed in experimental studies, of graphene at
large external magnetic fields.

\section{The magnetic field background on the lattice}

The introduction of a magnetic background
can be done defining on the lattice a gauge-invariant effective 
action $\Gamma(\vec{A}^{\text{ext}})$ by using the Schr\"{o}dinger 
Functional (SF)~\cite{Cea:1999gn}.
The Euclidean SF in Yang-Mills theories without matter is defined by
\beq
Z[A^{f},A^{i}]= \langle A^f | e^{-H T} \mathcal{P} | A^i \rangle \ , 
\eeq
{\it i.e.} it is the propagation kernel for going from some field 
configuration $A^i$, at time $x_4=0$, to some other configuration 
$A^f$, at $x_4=T$; the operator $\mathcal{P}$ projects onto the 
physical states.

The lattice SF is given by $Z[U^{f},U^{i}]=\int DU e^{-S}$, where $S$
is the Wilson action modified to take in account the boundaries:
$U(x)_{x_4=0}=U^i$ and $U(x)_{x_4=T}=U^f$.
We define the lattice effective action for a background 
field $\vec{A}^{\text{ext}}$ as
\beq
\Gamma(\vec{A}^{\text{ext}})= - \frac{1}{T} \ln{\left( \frac{\tilde{Z}
[U^{\text{ext}}]}{\tilde{Z}[0]}\right)}
\ , \quad \mbox{where }  \tilde{Z}[U^{\text{ext}}]=
Z[U^{\text{ext}},U^{\text{ext}}] \ .
\eeq
Remarkably, it turns out that $\Gamma(\vec{A}^{\text{ext}})$ is invariant 
under lattice gauge transformations of the external link $U^{\text{ext}}$.
Since in this definition $U^f=U^i$, we have periodic conditions in 
the time direction and, due to the lack of free boundaries, the lattice 
action is now the familiar Wilson action.

Moreover, it is possible to show that 
\beq
\Gamma(\vec{A}^{\text{ext}}) \to E_0(\vec{A}^{\text{ext}})-E_0(\vec{0}) \ , \quad 
(\mbox{when } T \to \infty ) \ ,
\eeq
where $E_0(\vec{A}^{\text{ext}})$  is the vacuum energy in presence of the 
external background field. 
Therefore $\Gamma(\vec{A}^{\text{ext}})$ is the lattice gauge-invariant 
effective action for the background field $\vec{A}^{\text{ext}}$.
In other words to study a theory with an external background field we 
have to simulate on the lattice the original action ({\it i.e.} the one 
without any external field), but introducing proper constraints.

In our case we are interested in $U(1)$ in a uniform external magnetic 
field $H$, therefore, after imposing spatial and temporal boundary conditions,
we have to constrain the spatial lattice links belonging to a fixed time slice
(for example $x_4=0$) to
\beq
U_1^{\text{ext}}(\vec{x})=1 \ , \quad U_2^{\text{ext}}(\vec{x})=\cos{(g H x_1)}+ i 
\sin{(g H x_1)} \ ,
\eeq
where $g$ is the coupling constant.
The same constraints are imposed at the spatial boundaries of the 
other time slices, {\it i.e.} we require that fluctuations over the background 
field vanish at infinity.
The temporal links are not constrained, which is in coherence with the 
correct definition of the thermal partition functional~\cite{Cea:2005td}.

We can see that, since the lattice has the topology of a torus, 
the magnetic field turns out to be quantized:
\beq
a^2 g H = \frac{2 \pi}{L_t} n_{\text{ext}} \ , \quad (n_{\text{ext}}= 0, 1, \dots) 
\ ,
\eeq
where $L_t$ is the temporal lattice extension.
A different approach to the problem we have tackled, where also a different way
to introduce the external magnetic field is implemented, can be 
found in Ref.~\cite{Alexandre:2001pa}.

This paper is based on the study of QED with $N_f=2$ flavours of 4-component 
fermions using the staggered fermion approach; this means that we need to 
simulate $N=1$ staggered fermions fields $\chi, \bar{\chi}$ with the Euclidean 
action~\cite{Fiore:2005ps}:
\beq
\label{action}
 S=S_G+\sum_{i=1}^{N} \sum_{n,k}  \overline\chi_i(n)
M_{n,k} \chi_i(k) \ ,
\eeq
and fermion matrix given by
\beq
M_{n,k}[U]=\sum_{\nu=1,2,3} \frac{\eta_{\nu}(n)}{2} \left\{ [U_{\nu}(n)
    ]  \delta_{k,n+\hat{\nu}}-
    [U_{\nu}^{\dagger}(k)] \delta_{k,n-\hat{\nu}} \right\} + m \ \delta_{n,k} \ ,
\eeq
where $\eta_{\nu}(n) = (-1)^{n_1+\ldots+n_{\nu-1}}$.
Moreover, we choose the compact formulation of QED,
\beq
S_G[U]= \beta \sum_{n,\mu < \nu}{ \left[ 1 - \frac{1}{2} \left( U_{\mu \nu}
 (n)+ U_{\mu \nu}^{\dagger}(n) \right) \right]} \ ,
\eeq
where $U_{\mu \nu}(n)$ is the ``plaquette variable'' and
$\beta=1/(g^2 a)$, $a$ being the lattice spacing.

Note that the introduction of the fermions in the theory does not 
change anything about the way we introduce the external field 
$\vec{A}^{\text{ext}}$~\cite{Cea:2004ux}.


\section{Dynamical symmetry breaking}

It is now a general result that a constant magnetic field leads to 
the generation of a fermion dynamical mass~\cite{Cea:1999jz}
in a wide class of $2+1$ and $3+1$ dimensional theories; 
this phenomenon is known as magnetic catalysis~\cite{Gusynin:1994re}.

In the QED in three dimensions it is possible to determine, at first order in 
perturbation theory, the value of the chiral condensate in presence of the 
magnetic field $H$ (in cgs units)~\cite{Cea:2011vi}:
\beq
\langle \overline\Psi \Psi \rangle \; = \; - \, 2 \, N_f \, |m| c^2 \, 
\frac { \hslash  c  eH} {2 \pi} 
\sum_{n=1}^{\infty} \frac {1} {\sqrt{2 n \hslash c eH+m^2 c^4}} \ .
\eeq
After regularisation of the integral and in the limit of small gap
$\Delta_0=m c^2$, we get:
\beq
\langle\overline\Psi\Psi\rangle  \; \simeq \;   \, - \,   
\frac{\hbar c eH}{2 \pi}  N_f \frac{\zeta(\frac{1}{2})}{\sqrt{\pi}}  
\frac{\Delta_0}{\sqrt{ \frac{\hslash c eH}{2\pi}}} \ .
\label{chicon}
\eeq
%


\section{Monte Carlo Simulations}

Our simulations have been performed in the weak coupling regime 
with $\beta=2.0$, on volumes $L^3$, with $L=12, 16, 24$,  
for masses in the range $0.005 < m_0 < 0.05$ and for several
values of the strengths of the magnetic field $n_{\text{ext}}$,
namely, $n_{\text{ext}}=1,2,3$. The Monte Carlo simulation code 
is based on the Hybrid Monte Carlo algorithm.

\begin{figure}[ht]
\center
\vspace{10mm}
\includegraphics[width=15cm]{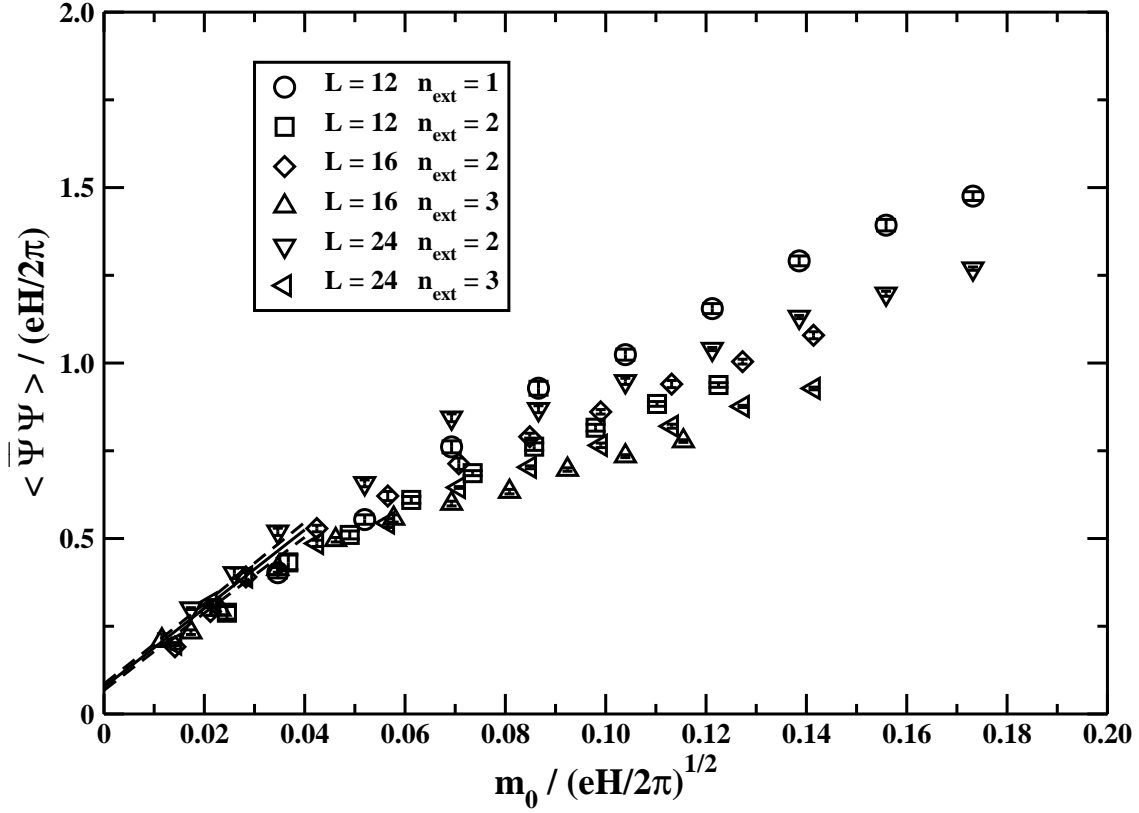}
\caption{Scaled chiral condensate versus the scaling variable $x$. 
The continuum line is the linear fit of the data in the scaling region 
$0 < x \lesssim 0.04$.}
\label{plot1}
\end{figure}

From Eq.~(\ref{chicon}) we can point out the natural dimensionless quantities
to be used in the plot of our results:
\beq
y=\frac{ \langle\overline\Psi\Psi\rangle }{\frac{eH}{2\pi}} 
\quad
\mbox{and}
\quad
x=\frac{m_0}{\sqrt{\frac{eH}{2 \pi}}} \ .
\label{variablexy}
\eeq
In Fig.~\ref{plot1} we plot our data all together, regardless of possible 
finite volume effects that are nevertheless smaller than statistical errors.
We see that the continuum scaling law is satisfied for $0 < x \lesssim 0.04$.
This allows us to extract the chiral condensate in the chiral limit, 
$x \rightarrow 0$, by the following linear relation:
\beq
\label{eqfit}
y \; = \; a_0 \,+ \, a_1 \, x \ .
\eeq
The best fit of the data to Eq.~(\ref{eqfit}), in the scaling 
region, gives
\begin{equation}
\label{fitresults}
 a_0 \, = \, 0.07668 \, \pm \, 0.00930 \;  \; , \qquad a_1 \, = \, 11.20 \, 
\pm \, 0.48 \ .
\end{equation}
As a consequence, we see that in the chiral limit the external magnetic 
field does induce a non-zero chiral condensate. 
From Eq.~(\ref{variablexy}) and the result in Eq.~(\ref{fitresults}), 
restoring cgs units, we find:
\begin{equation}
\label{chiconnum}
\langle\overline\Psi\Psi\rangle  \; = \;  \frac{\hbar c eH}{2 \pi}  
\left(  0.07668 \, \pm \, 0.00930 \right) \, .
\end{equation}
The non-zero value of the chiral condensate can be interpreted as 
the generation of a dynamical fermion mass $m_{\text{dyn}}$ with gap  
$\Delta_0 = m_{\text{dyn}} c^2$.
To extract $\Delta_0$  we use Eq.~(\ref{chicon}) combined with 
Eq.~(\ref{chiconnum}):
\begin{equation}
\label{deltanum}
\Delta_0  \; \simeq \;   \, - \,  \frac{\sqrt{\pi}}{  N_f \zeta(\frac{1}{2})}
\sqrt{ \frac{\hslash c eH}{2\pi}}   \left(  0.07668 \, \pm \, 0.00930 
\right) \, .
\end{equation}
%


\section{Application to graphene}

The graphene is a single sheet of carbon atoms characterised by
a honeycomb lattice described by two interpenetrating triangular 
Bravais lattices. The unit cell is instead a rhombus containing two 
carbon atoms. The reciprocal lattice comes out to be a hexagonal
as well but, between the six corners of the Brillouin zone, only two
points are inequivalent ({\it i.e.} not connected by a reciprocal lattice 
vector) and are labelled by $K$ and $K^\prime$, called ``valleys''.

A simple tight-binding model~\cite{wallace} shows that the dispersion 
relation $E(\mathbf{k})$ goes to zero at $K$ and $K^\prime$.
Everywhere else in momentum space $E(\mathbf{k}) \neq 0$ and we have 
two graphene bands, conduction and valence, lying symmetrically above and 
below the Fermi energy at $E = 0$.
Therefore, the Fermi level is placed between the two symmetrical bands, 
with zero excitation energy needed to excite an electron from just below 
the Fermi energy to just above the six points of the corners of the 
Brillouin zone.

Moreover, the dispersion relation turns out to be linear near the 
$K$ and $K^\prime$ points: $E(\mathbf{k})= \pm \hslash v_F | \mathbf{k} |$,
where $v_F$ is the Fermi velocity.
As discussed in Ref.~\cite{semenoff}, this yields an effective Hamiltonian 
made of two copies of Pauli spinors which satisfy the massless two-dimensional 
Dirac equation with the speed of light replaced by the Fermi velocity.

When graphene is immersed in a transverse magnetic field $H$ the 
energy levels are quantised into non-equidistant Landau levels:
\beq
E_n= \mbox{sign}(n) \sqrt{2 \hslash e H |n|\frac{v_F^2} {c}},
\quad
n=0, \pm 1, \ldots \ .
\eeq
The presence of the ``anomalous'' Landau level at zero energy leads to 
half-integer 
Quantum Hall (QH) effect; the Hall resistance is
given by $R_{xy}=h/(e^2 \nu)$, where the quantised filling factor $\nu$ is: 
\beq
\nu=\pm g_s (N+\frac{1}{2})=\pm 2, \pm 6, \pm 10, \dots \ .
\eeq
The factor $g_s=4$ takes into account the spin and valley degeneracy
and $\pm$ stands for electron and holes, respectively.

Recently, it has been shown~\cite{Zhang:2006aa} that for very 
strong magnetic field ($H \gtrsim 20 \, T$) new QH states appear 
corresponding to $\nu=0, \pm 1, \pm4$.
This means that the $n=0$ Landau level is totally resolved into $\nu=0, \pm1$
plateaus, while the fourfold degeneracy in the $n=\pm 1$ Landau levels 
is only partially resolved into $\nu=\pm 4$ leaving a 
twofold degeneracy~\cite{Zhang:2006aa}. 

The new plateaus at $\nu=0, \pm4$ can be explained by Zeeman spin 
splitting; $\nu=\pm 1$ can be explained if there is some kind of
generation of a gap $\Delta_0$, {\it i.e.} a valley symmetry breaking 
in $n=0$ Landau level~\cite{Zhang:2006aa,Jiang:2007aa}.

In Ref.~\cite{Jiang:2007aa} the authors determine experimentally the
$\Delta E (\nu = 1)$ dependence from the magnetic field: they find
for it a $\sqrt{H}$ behaviour. 

We have fitted the experimental data from Ref.~\cite{Jiang:2007aa} to
\beq
\label{eq1.4}
\Delta E (\nu = 1)  =  2 \, \left ( \Delta_0(H) \, - \,  \frac{g}{2} 
\mu_B \, H \right ) \ ,
\eeq
where $\mu_B$ is the Bohr magneton,  $g=2$ and $\Delta_0(H) \sim \sqrt{H}$ 
(note that the second term takes into account the Zeeman effect);  we find:
\beq
\label{eq_delta_exp}
\Delta_0(H) \, = \, (13.57 \, \pm 0.28) \; {\rm K} \times k_B \; \sqrt{H(T)} \ ,
\eeq
where $k_B$ is the Boltzmann factor and $H(T)$ stands for the magnetic field
expressed in Tesla.

It is believed that the generation of the gap is driven by the 
electron-electron interaction (for a review see Ref.~\cite{Kotov:2010yh});
in this picture, the value of the gap is expected to be of the order:
\beq
\Delta_0(H) \approx \frac{e^2}{\epsilon} \sqrt{\frac{eH}{\hslash c}}
\approx 163 K \times k_B \sqrt{H(T)} \ ,
\eeq
{\it i.e.} this result is one order of magnitude bigger than the 
experimental result of Eq.~(\ref{eq_delta_exp}).

On the other hand, in Ref.~\cite{Cea:2011vi} it is supported that the 
origin of the gap
is related to a \emph{dynamical} generation related to the presence of the 
magnetic field; this means that we can derive the value of the gap 
by our numerical results.

The implied hypothesis here is that the Coulomb interaction between
the electrons can be neglected because the dominant effect is the
magnetic catalysis; this is the reason why we can apply planar QED directly
to describe the graphene, although in the literature the Coulomb interaction
is always considered as a $3d$ field acting on $2d$ fermions.

To apply our results to graphene, we must simply replace the speed 
of light $c$ with the Fermi velocity $v_F$ in  
Eq.~(\ref{deltanum})~\cite{Cea:2011vi}:
\begin{equation}
\label{eqdeltapart}
\Delta_0  \; \simeq \;   \, - \,  \frac{\sqrt{\pi}}{  N_f \zeta(\frac{1}{2})}
\sqrt{ \frac{\hslash \frac{v_F^2}{c} eH}{2\pi}}   \left(  0.07668 \, 
\pm \, 0.00930 \right) \ .
\end{equation}

It is worth mentioning that the small gap approximation, which we are 
using here, is correct when it is satisfied the following relation:
\beq 
\frac{\Delta_0}{\sqrt{H(T)}} \ll \sqrt{2 \hslash v_F^2 e /c} 
\approx 420 K \times k_B \ ;
\label{deltamin}
\eeq
the experimental value found in
Eq.~(\ref{eq_delta_exp}) is actually small compared with that in 
Eq.~(\ref{deltamin}), therefore we can consider the small gap 
approximation valid in this context.

Finally, using the experimental value for the Fermi velocity 
($v_F \approx 1.0 \times 10^8$cm/s) in Eq.~(\ref{eqdeltapart}), we get :
\begin{equation}
\label{eq2.15}
\Delta_0(H) \, = \, (5.52 \, \pm 0.67) \; {\rm K} \times k_B \; \sqrt{H(T)} \ . 
\end{equation}
This result can be compared directly with Eq.~(\ref{eq_delta_exp}). 
Since the two values are of the same order of magnitude, 
our result also confirm our hypothesis that the Coulomb interaction
between electrons is actually negligible in presence of an 
external magnetic field.

\section{Conclusion}

Our nonperturbative Monte Carlo simulations have shown that the external 
magnetic field gives rise to a spontaneous breaking of the chiral symmetry.
We think that it is possible to apply our results to describe the breaking 
of the valley symmetry in graphene under strong magnetic field and we
determine the numerical value of the gap that compares quite 
well with the experimental value.


\end{document}